\documentclass[]{IEEEtran}

\usepackage{color}
\usepackage{cite}
\usepackage[colorlinks=true,linkcolor=blue,urlcolor=blue,citecolor=blue]{hyperref}
\usepackage{upgreek}
\usepackage{footmisc}

\usepackage[colorinlistoftodos]{todonotes}

\begin{document}

\IEEEpubid{\makebox[\columnwidth]{979-8-3315-4683-0/26/\$31.00~\copyright2026 IEEE \hfill} \hspace{\columnsep}\makebox[\columnwidth]{ }}

\title{\LARGE Recent Advances in Tabletop Kibble Balance -- KBmini}

\author{Nanjia Li, Weibo Liu, Kang Ma, Elsayed E.E. Qupasie, Wei Zhao, Songling Huang, Shisong Li$^\dagger$\\
Department of Electrical Engineering, Tsinghua University, Beijing 100084, China\\
$^\dagger$Email: shisongli@tsinghua.edu.cn}

\maketitle

\IEEEpubidadjcol

\begin{abstract}

This paper presents recent advances in the KBmini Kibble balance, a tabletop system for E2-accuracy mass calibration up to 1\,kg. The $Bl(z)$ profile is characterized by manually setting the magnet at different vertical positions, and the extremum point is selected as the weighing position. The spring constant of the weighing cell around this point is measured. With a new coil of a larger number of turns and a multi-harmonic excitation technique, a near-constant velocity profile over a moving range of 180\,$\upmu$m, producing an induced-voltage flat-top region exceeding 1\,V, is achieved. These results establish a foundation for subsequent mass calibration experiments.
\end{abstract}

\begin{IEEEkeywords}
Kibble balance, mass calibration, harmonic excitation, magnetic measurement.
\end{IEEEkeywords}

\pagenumbering{gobble}

\section{Introduction}

Following the implementation of the 2019 new definition of the kilogram, the Kibble balance has emerged as one of the primary techniques for realizing a mass with direct traceability to the Planck constant \(h\). Traditional Kibble balances are large, complex systems typically confined to national metrology institutes (NMIs)~\cite{NIST2017,NRC2017}. To extend this quantum-traceable calibration capability to industrial and wider laboratory settings, the development of compact, tabletop Kibble balances has become a key research frontier, e.g.~\cite{NISTtabletop2020,NPLtabletop2024,PTB_PB2_2021,THU2025}. Following a tabletop setup for NMI-level mass realization at Tsinghua University~\cite{THUdesign2022,THUKB2023}, a new tabletop Kibble balance system, KBmini, targeting E2-accuracy mass calibrations from gram-level to 1\,kg, has been launched in 2025. Based on primary conception demonstrations~\cite{KBmini}, this paper presents recent experimental progress on the KBmini system. Operated in air, the KBmini system now includes updated measurements of $Bl$ distribution, flexure hinge spring constant, and multi-harmonic excitation.

\section{Overall design}

\begin{figure}[tp!]
\centering
\includegraphics[width=0.45\textwidth]{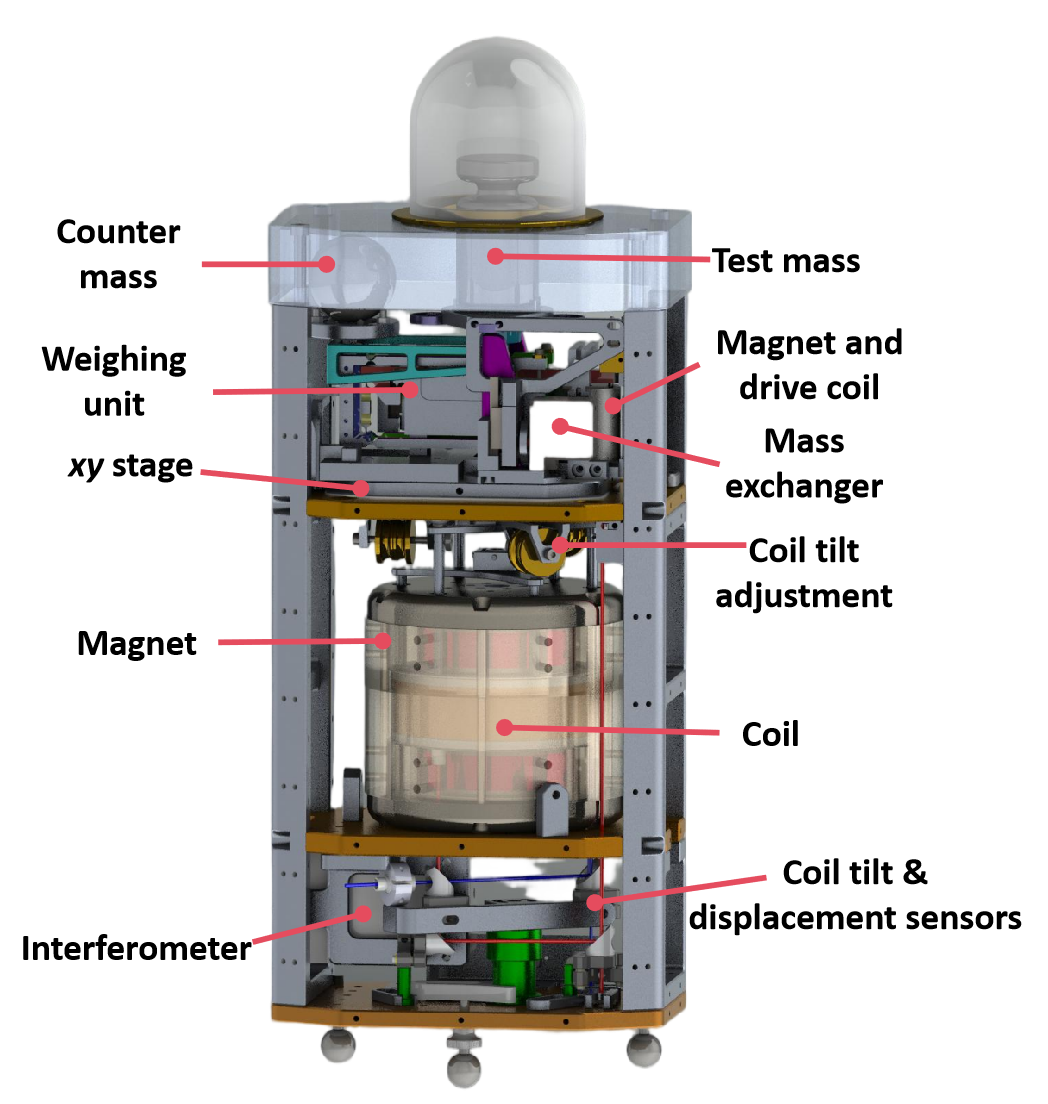}
\caption{CAD model of the KBmini. }
\label{KBmini}
\end{figure}

The core components of the KBmini include the weighing unit, mass exchanger, magnet system, coil, $xy$ adjustment stage, coil tilt adjustment, interferometer, motion sensors, and electrical references, as illustrated in Fig.\ref{KBmini}. The parameters and setup of KBmini have been presented in \cite{KBmini}.
A new coil, with the number of turns increased to 12 500 using 0.1\,mm diameter wire, 30\,mm in height and an average radius of 54.9\,mm, is integrated into the experiment. Its resistance is approximately 9.85 k$\Omega$. This design ensures that the operating current remains within 2\,mA, fully utilizing the 2\,mA range of the current source (Keithley 6221). The high $Bl$ value, in the meanwhile, allows an induced voltage exceeding 1\,V in the moving phase.

\section{Experimental progress}

\subsection{\textit{Bl} Measurement }

\begin{figure}[tp!]
\centering
\includegraphics[width=0.45\textwidth]{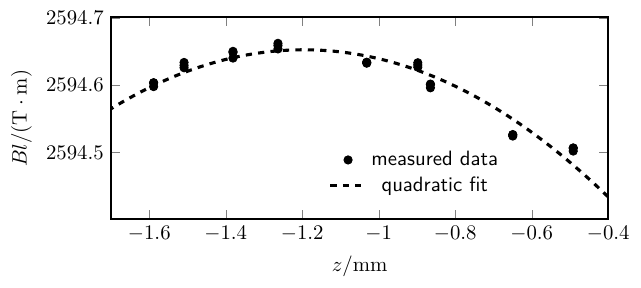}
\caption{Measurement result of $Bl(z)$ profile.}
\label{Bl}
\end{figure}

The $Bl(z)$ profile was determined through weighing measurements conducted by moving the magnet to various vertical positions $z$. The origin of the $z$-axis is defined at the center of the magnet, with the positive direction pointing upward.
The $Bl$ value at each position is derived from the equation \( Bl = mg / (I_+ - I_-) \), where $I_+$ and $I_-$ are the coil currents measured with and without a 1\,kg stainless steel mass, respectively.
The measurement results are presented in Fig.\ref{Bl}. The markers indicate the experimental data, while the dashed line represents a quadratic fit. Over the measured range, $Bl(z)$ shows a clear quadratic dependence on $z$, with its extremum occurring at $z = -1.20$\,mm. Due to the relatively flat variation of $Bl$ near this extremum, the weighing position was set at $z = -1.20$\,mm.

\subsection{Spring Constant Measurement}

\begin{figure}[tp!]
\centering
\includegraphics[width=0.45\textwidth]{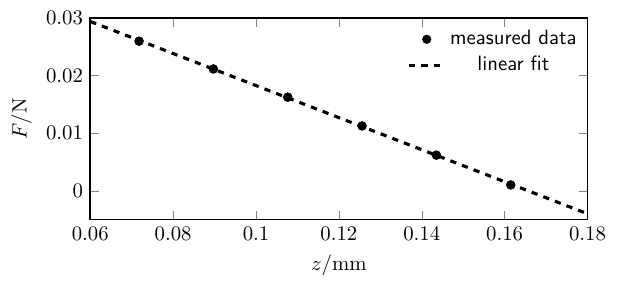}
\caption{$F(z)$ curve of the weighing unit. }
\label{spring constant}
\end{figure}

By varying \(z\) while maintaining balance via PID control, the magnetic force $F$ as a function of $z$ can be measured. The resulting $F(z)$ curve is shown in Fig.\ref{spring constant}. The linear fit indicates a good linearity, yielding a spring constant of the weighing unit \(k = 277.8\,\mathrm{N/m}\).

\subsection{Multi-harmonic Excitation Measurement}

\begin{figure}[tp!]
\centering
\includegraphics[width=0.45\textwidth]{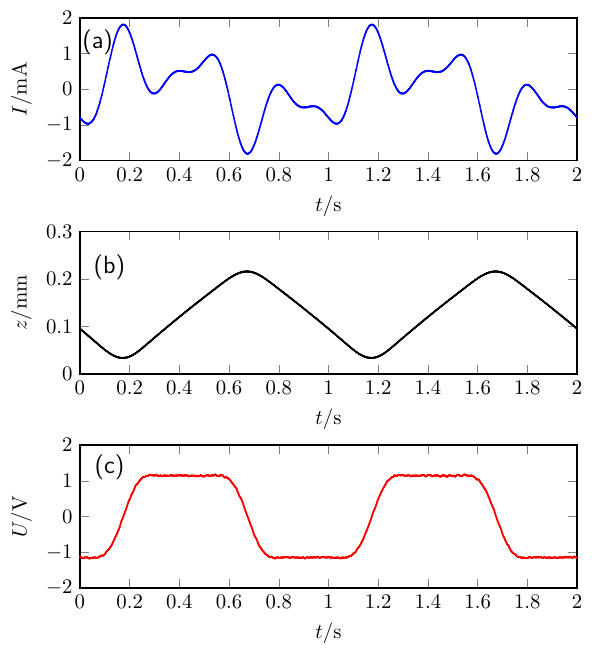}
\caption{Multi-harmonic excitation results for the new coil: (a) drive current $I$, (b) coil displacement $z$, and (c) induced voltage $U$ as functions of time.} 
\label{multi-harmonic excitation}
\end{figure}

The multi-harmonic excitation method \cite{KBmini} is employed to generate a near-constant velocity profile by injecting odd-order harmonic currents into the drive coil.
For the new coil, the optimized drive current consists of the fundamental frequency and its odd harmonics. Through iterative optimization, the amplitude ratios were determined as \(A_\mathrm{1} : A_\mathrm{3} : A_\mathrm{5} : A_\mathrm{7} = 0.72 : 0.80 : 0.42 : 0.10\), with corresponding phases \(\phi_\mathrm{3} = -0.27\,\mathrm{rad}\), \(\phi_\mathrm{5} = -1.30\,\mathrm{rad}\), and \(\phi_\mathrm{7} = -2.70\,\mathrm{rad}\). The peak current in the drive coil during the cycle is 1.8\,mA. Under this excitation, the induced voltage exhibits a flat-top region exceeding 1\,V, which occupies 68\% of the full oscillation cycle ($\Delta U/U<$ 5\%). The induced voltage above 1\,V facilitates differential measurements against a Zener reference. Fig.\ref{multi-harmonic excitation} presents the measured time traces of the drive current, coil displacement, and induced voltage under this setup.

\section{Conclusion}

This paper presents recent advances in the KBmini Kibble balance, including $Bl(z)$ profile characterization, weighing cell spring constant evaluation, and optimization of the multi-harmonic excitation.
From the $Bl(z)$ measurement, an extremum at $z = -1.20$\,mm is selected as the weighing position. The spring constant in the vicinity is determined as 277.8~N/m.
A multi-harmonic excitation method achieves a near-constant velocity profile for the new coil. Under optimized excitation, the induced voltage exhibits a flat-top region exceeding 1~V, occupying 68\% of the oscillation cycle with fluctuation below 5\%. The present experimental progress helps to demonstrate the forthcoming weighing and velocity measurements. 
\section*{Acknowledgment}

This work was supported by the National Natural Science
Foundation of China under Grant 52377011.

\bibliographystyle{unsrt}

\end{document}